\documentclass[journal=nalefd,manuscript=article]{achemso}
\setkeys{acs}{articletitle = true} 
\usepackage{graphicx}
\usepackage{dcolumn}
\usepackage{bm}
\graphicspath{ {figures/} }
\usepackage{hyperref}

\title{Approaching Unity Photon Collection from NV Centers via  Ultra-Precise Positioning of Nanodiamonds  in Hybrid Nanoantennas}

\author{Boaz Lubotzky}
\affiliation{Racah Institute of Physics, The Hebrew University of Jerusalem, Jerusalem 9190401, Israel}
\affiliation{The Center for Nanoscience and Nanotechnology, The Hebrew University of Jerusalem, Jerusalem 9190401, Israel}
\author{Hamza Abudayyeh}
\affiliation{Racah Institute of Physics, The Hebrew University of Jerusalem, Jerusalem 9190401, Israel}
\affiliation{The Center for Nanoscience and Nanotechnology, The Hebrew University of Jerusalem, Jerusalem 9190401, Israel}

\author{Niko Nikolay}%
\affiliation{AG Nanooptik, Humboldt Universitat zu Berlin, Newtonstraße 15, D-12489 Berlin, Germany}
\affiliation{IRIS Adlershof, Humboldt Universitat zu Berlin, Zum Großen Windkanal 6, 12489 Berlin, Germany}
\author{Oliver Benson}%
\affiliation{AG Nanooptik, Humboldt Universitat zu Berlin, Newtonstraße 15, D-12489 Berlin, Germany}
\affiliation{IRIS Adlershof, Humboldt Universitat zu Berlin, Zum Großen Windkanal 6, 12489 Berlin, Germany}
\author{Ronen Rapaport}
\affiliation{Racah Institute of Physics, The Hebrew University of Jerusalem, Jerusalem 9190401, Israel}
\affiliation{The Center for Nanoscience and Nanotechnology, The Hebrew University of Jerusalem, Jerusalem 9190401, Israel}
\email{ronen.rapaport@huji.ac.il}

\begin{document}

\begin{abstract}
Efficient readout of nitrogen-vacancy (NV) centers in diamond is crucial for various quantum information technologies. However, achieving high-fidelity, single-shot readout at room temperature remains challenging due to limited photon collection efficiency (CE) and background noise. In this work, we enhance the readout efficiency of NV centers by integrating them into hybrid metal-dielectric bullseye nanoantennas using ultra-precise deterministic positioning. This approach enables highly directional emission while minimizing optical losses, resulting in a measured CE of 82\% into a numerical aperture (NA) as low as 0.5, and approaching unity for NA$>$0.8. This marks a substantial improvement over previous realizations using nanodiamonds, highlighting the advantage of combining hybrid nanoantennas with precise positioning. Our results mark a substantial advancement towards efficient single-shot readout of NV centers by significantly improving readout fidelity and efficiency in a simple on-chip configuration.
\end{abstract}

\maketitle

\begin{figure}[t] 
 \centering
 \includegraphics[width=80mm,scale=1]{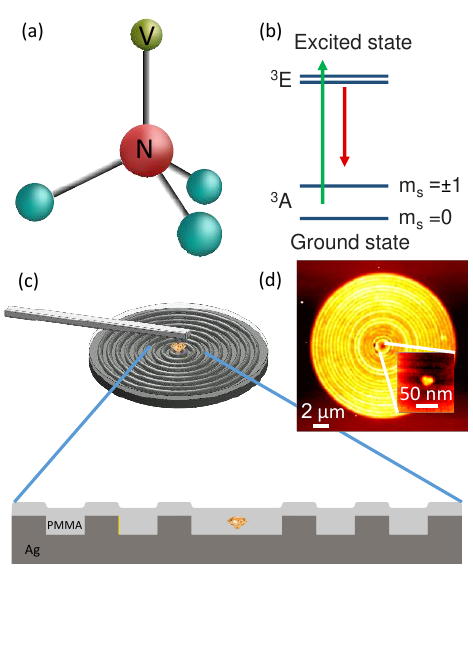}
 \caption{(a) NV center atomic structure in diamond, showing the nitrogen atom, the vacancy, and surrounding carbon atoms. (b) Simplified energy level diagram of the NV center, illustrating the ground state (\(^3A\)) with separated \(m_s = 0\) and higher energy \(m_s = \pm1\) levels, and the excited state (\(^3E\)) showing two closely spaced levels. The green arrow represents optical excitation from the ground state to the excited state, while the red arrow represents photon emission during the transition from the excited state back to the ground state. (c) Schematic representation of the ND placement approach and the bullseye antenna. (d) AFM scans of the placed ND in the center of the bullseye antenna.
 \label{Fig: method} }
\end{figure}

Efficient readout of the spin state of an NV center in diamond is a fundamental challenge in the development of quantum technologies  based on NV centers \cite{Wolf2015Purcell-enhancedDiamond}. NV centers are point defects in the diamond lattice, characterized by a substitutional nitrogen atom adjacent to a lattice vacancy, as depicted in Figure \ref{Fig: method}(a). The distinct atomic structure and associated energy levels of an NV center yield a unique combined spin and optical properties, including long spin coherence times as well as spin-dependent fluorescence, as illustrated in Figure \ref{Fig: method}(b). This combination  makes  the NV platform a promising candidate for quantum metrology and information processing \cite{Doherty2013, Guo2023}.

However, the emission from the optical transitions of an NV center is inherently non-directional, leading to low photon collection efficiency (CE) and restricting its usability in high-fidelity readout applications. Enhancing CE for low NA optical elements is essential to improve readout fidelity and enable scalable integration of NV centers in quantum devices \cite{Hausmann2012, Yufan2015,Dhomkar2024}. While previous works have explored various strategies, including dielectric antennas and plasmonic structures, achieving near-unity CE at room temperature remains a challenge due to optical losses and misalignment issues \cite{Robledo2011, Hensen2015, Hopper2018, Zhang2021}.  

In this work, we employ a hybrid metal-dielectric nanoantenna combined with ultra-precise deterministic positioning to significantly enhance photon collection from NV centers in nanodiamonds. This approach achieves a measured CE of 82\% for NA = 0.5 and CE that approaches unity for NA$>$0.8, demonstrating a substantial improvement in photon extraction efficiency.

Various near-field coupling configurations have been explored to enhance the directionality of quantum emitters. Metal-based nanoantennas \cite{Nepal2013LargeFluorescence, Hoang2015UltrafastNanoantennas, Ahmed2012PhotoluminescenceNanoneedles, Guo2015ControllingNanoarrays, Bitton2019QuantumCoupling} provide broadband enhancement due to their small mode volumes and low-quality factors, but suffer from significant ohmic losses \cite{Bitton2019QuantumCoupling}. Dielectric nanoantennas, while highly directional and low-loss, are inherently limited to narrow-band emitters and often experience strong emission into the high-index dielectric substrate, reducing practical collection efficiency \cite{Skolnick1998StrongStructures, Press2007PhotonRegime, Bogdanovic2017DesignNetworks, Jun2011PlasmonicEmission}.

A hybrid metal-dielectric bullseye nanoantenna offers a unique solution by combining the strengths of both metallic and dielectric structures, achieving high directionality and near-unity CE over a broad spectral range at room temperature \cite{Livneh2015EfficientNanoantenna, livneh2016highly, abudayyeh2017quantum, Abudayyeh2021QD_Placement, Lubotzky2024, Waltrich2021}. The metallic structure enables efficient diffraction over a broad spectral range, enhancing directionality, while the dielectric waveguide layer facilitates photon extraction with minimal nonradiative losses. By integrating NV-containing nanodiamonds into such nanoantennas with ultra-precise positioning, we achieve optimal coupling, resulting in record-high photon extraction efficiencies.

The hybrid metal-dielectric bullseye nanoantenna configuration allows for precise control over the emission pattern, directing the photons into a narrow angular cone, thereby enhancing overall emission efficiency. The rotational symmetry of the bullseye nanoantenna is particularly attractive as it effectively directs the emission of dipole emitters regardless of their XY orientation, ensuring high directivity \cite{abudayyeh2017quantum}. This dipole orientation insensitive directivity is crucial for  NV-based quantum  applications requiring high CE, making  such a system well-suited for integration with optical fibers and other photonic devices \cite{Lubotzky2024}. 
We therefore utilize here ND integrated with such a hybrid metal-dielectric bullseye nanoantenna.

Effective coupling between a quantum emitter and a nanoantenna requires a precise positioning method. Several techniques have been explored for different antenna configurations, such as near-field optical lithography, scanning probe manipulation, and nanowire-based techniques \cite{Sipahigil2016AnNetworks, Gschrey2013InSpectroscopy, Shi2016WiringLithography, Harats2017DesignEmission, Schell2011ADevices, Cuche2009DiamondOptics, VanDerSar2009NanopositioningCenter, Huck2011ControlledNanowire}. In this study, we utilized a scanning probe-based pick-and-place method, as detailed in the work by Nikolay et al. \cite{Nikolay2018AccurateStructures}, to deterministically and reproducibly place the NDs onto the nanoantennas with nanometric precision. This technique is both fast and versatile, allowing for high-accuracy placement even on opaque structures. It also enables precharacterization of the emitter, facilitating the selection of suitable emitters for specific applications. Moreover, this method allows for iterative adjustments to the nanoparticle's position, ensuring deterministic coupling and optimal device performance.

We selected NDs containing NV centers, with a focus on those that house ensembles of NV centers, which are known to enhance sensitivity in sensing applications \cite{Webb2020OptimizationSignals, Taylor2008High-sensitivityResolution}. Additionally, NDs with single NV centers were chosen for their quantum optical characterization. These NDs were then placed on dielectric polymethylmethacrylate (PMMA) coated Ag bullseye antennas developed and detailed in Refs. \cite{Harats2017, Abudayyeh2021QD_Placement, livneh2016highly}, as shown schematically in Fig \ref{Fig: method}. An Atomic force microscopy (AFM) scan confirming the successful placement of the ND in the center of the bullseye antenna is shown in Figure \ref{Fig: method}(d).

After the precise placement of the ND, another PMMA layer is spin-coated on top of the hybrid nanoantenna, forming a complete waveguide layer that contributes to the hybrid metal-dielectric nanostructure. This additional layer enhances the collimation of the emission from the ND, as described in \cite{Abudayyeh2021QD_Placement}. The final configuration, including the ND embedded in the device, is depicted in Figure \ref{Fig: method}(c), which provides a schematic representation of the ND placement approach and the bullseye antenna structure.

\begin{figure}[t] 
 \centering
 \includegraphics[width=160mm,scale=1]{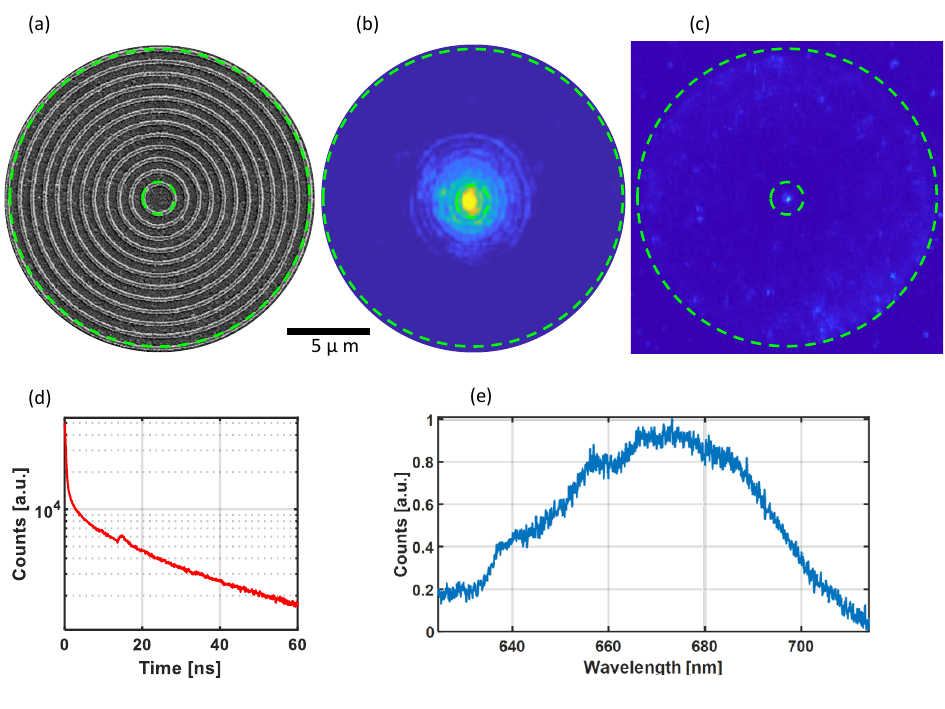}
 \caption{Results from a bullseye nanoantenna coupled to a single ND, which contains one or more NV centers. The scale bar refers to (a)-(c). The dashed green lines in (a)-(c) mark the first and last circular gratings of the bullseye structure, providing a visual guide to the extent of the nanoantenna. (a) A SEM image of the bullseye nanoantenna before the deposition of the dielectric layers and the positioning of the ND. (b) Spatial PL map of the emitter device obtained using wide-field imaging. (c) Confocal scan of the antenna with the NV center in the middle. A pulsed laser was used and the counts during the first 2 nanoseconds after each laser pulse were omitted to implement temporal filtering for  suppression of background fluorescence.. (d) NV center fluorescence lifetime measurement on a semi-logarithmic scale. (e) NV center emission spectrum measurement.}
 \label{Fig: Single ND}
\end{figure}

In Figure \ref{Fig: Single ND}, we present key results from our bullseye nanoantenna coupled with a single ND. Figure \ref{Fig: Single ND}(a) shows an scanning electron microscope (SEM) image of the fabricated metal part of the bullseye nanoantenna before the deposition of dielectric layers and the positioning of the ND. This image provides a reference for the structural features of the nanoantenna.

Figures \ref{Fig: Single ND}(b) and \ref{Fig: Single ND}(c) display a spatial photoluminescence (PL) map and a confocal scan of the antenna with the ND at its center, respectively. These measurements were performed with a single nanodiamond (ND), which contains one or more NV centers.
 Figure \ref{Fig: Single ND}(b) was obtained using a tightly focused 532 nm continuous-wave (CW) diode laser for excitation, while Figure \ref{Fig: Single ND}(c) was obtained using a pulsed laser with a repetition rate of 5 MHz, omitting counts during the first 2 nanoseconds after each laser pulse to reduce background noise. The NV center emission was filtered using a combination of a 620 nm short-pass and a 720 nm long-pass filter and collected using an objective lens with a NA of 0.9. The consistent scaling and centering across these images, along with the SEM image in Figure \ref{Fig: Single ND}(a), allow for a direct visual correlation between the physical layout of the nanoantenna and the observed PL results, effectively illustrating the coupling effects achieved. 

The dashed green lines in Figure \ref{Fig: Single ND}(a) mark the first and last circular gratings of the bullseye structure, providing a visual guide to the extent of the nanoantenna. This marking is crucial for understanding the spatial confinement and the role of the nanoantenna in directing the emission, as also demonstrated in similar studies such as those by Abudayyeh et al. on deterministic placement of quantum dots in nanoantennas \cite{Abudayyeh2021QD_Placement} and by Nikolay et al. on the accurate placement of single nanoparticles on opaque conductive structures \cite{Nikolay2018AccurateStructures}. The images confirm the findings that fluorescence is primarily emitted from the central region of the bullseye structure, highlighting the precise positioning of the ND and the efficient coupling achieved in the device.

Figures \ref{Fig: Single ND}(d) and \ref{Fig: Single ND}(e) present measurements of fluorescence lifetime and emission spectrum, respectively. These measurements, performed using a 532 nm pulsed laser, reveal the characteristic NV emission spectrum at room temperature. The emission spectrum, shown in Figure \ref{Fig: Single ND}(e), later provides the weightings for different emission wavelengths in the simulation of the back focal plane image, shown in Figure \ref{Fig: PL}(d). 
This precise mapping of emission is essential for optimizing device performance, particularly in quantum applications that rely on controlled photon emission and high CE. 

\begin{figure}[t] 
 \centering
 \includegraphics[width=160mm,scale=1]{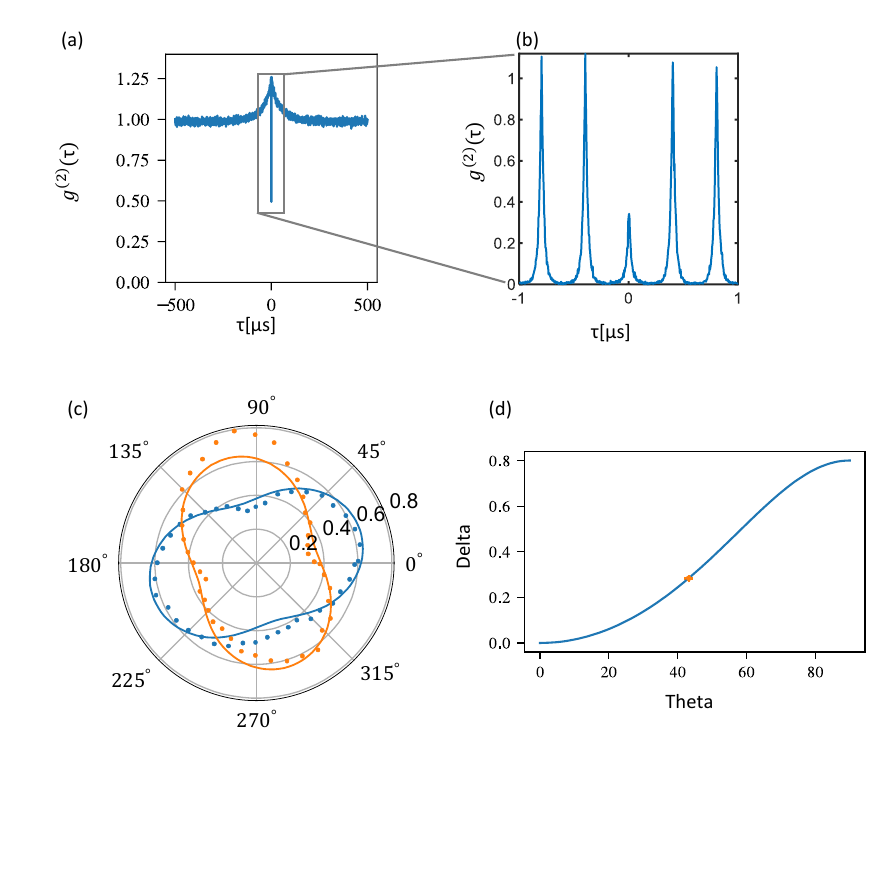}
 \caption{Results from a ND containing single NV center coupled to a bullseye nanoantenna. (a) A photon autocorrelation ($g^{(2)}(\tau)$ function) presenting a bunching at finite delay. (b) A photon autocorrelation ($g^{(2)}(\tau)$ function) showing an antibunching of less than 0.5, indicating single-photon emission. The zero-phonon line of NV centers at 637 nm is marked with a dashed green line. (c) Polar plot of intensity as a function of the polarization analysis angle for two perpendicular polarizations. (d) A map of degree of polarization and expected angle (theta) of the NV center crossed linear dipoles.}
 \label{Fig: Single NV}
\end{figure}

\begin{figure}[t] 
 \centering
 \includegraphics[width=160mm,scale=1]{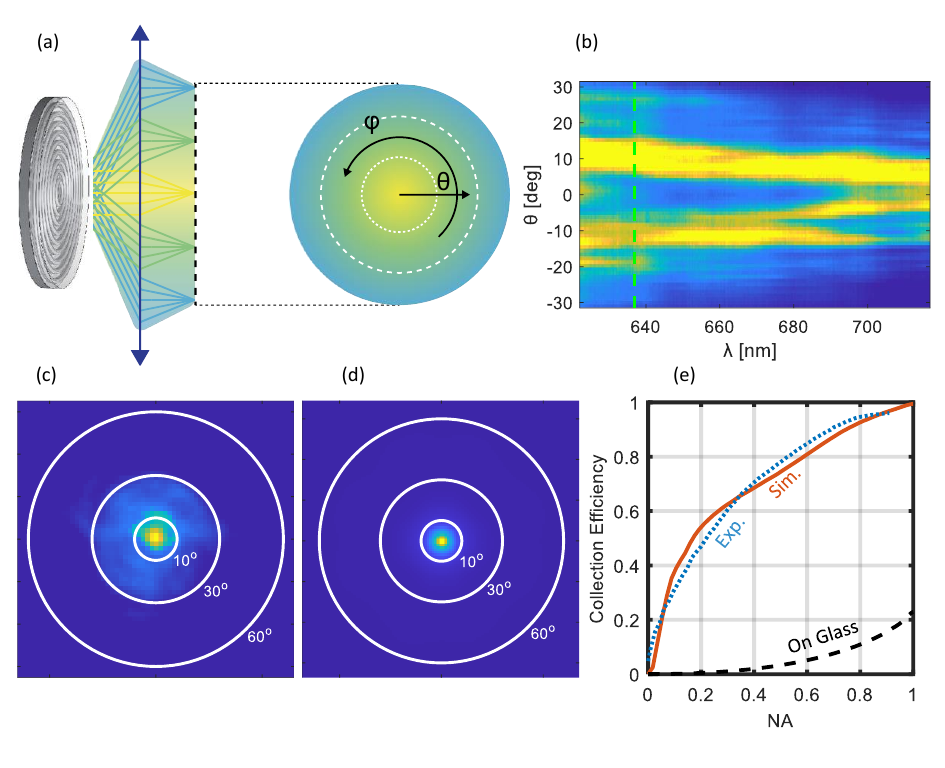}
 \caption{Back focal plane and collection efficiencies measurements and calculations of the device. (a) Schematic representation of the back focal plane imaging technique used to measure directionality in this study. (b) Spectrally resolved back focal plane PL image of a final device. The image is normalized with respect to the spectrum, i.e., each column (wavelength) is normalized to the max in that line. The zero-phonon line of NV centers at 637 nm is marked with a dashed green line. (c) Measured back focal plane image of the device. (d) Finite-difference time-domain (FDTD) simulation of the back focal plane image of the device. The simulation is a weighted average of the back focal plane of the device for different wavelengths, where the weights are obtained from the measurement of the spectrum of the NV as shown in Figure \ref{Fig: Single ND}(e). The simulation also takes into account the dipole orientation of the NV center, as described in Figure \ref{Fig: Single NV}. (e) Measured (blue) and simulated (red) CE of the device emission. Black: finite-difference time-domain (FDTD) simulation of a reference NV placed on a glass slide.
 \label{Fig: PL} } 
\end{figure}

Figure \ref{Fig: Single NV}(a) shows second order correlation measurement $g^{(2)}(\tau)$ for a time scale longer than the lifetime of the NV center emission, using pulsed laser at a repetition rate of 2.5 MHz. Figure \ref{Fig: Single NV}(b)  displays a zoom into short delay times, showing a clear antibunching behavior which confirms that indeed there is only single NV center emitting in this device. The shelving in a metastable singlet state (not shown in Fig. \ref{Fig: method}(b)) leads to the observation of photon bunching with longer timescales \cite{Thiering2018}. Following \cite{Lethiec2014}, we used a half-wave plate and a polarizing beam-splitter to separate the PL into two arms. Each arm leads to a single-photon counting avalanche photodiode. By rotating the half-wave plate we measured simultaneously the x- and y-polarized emission of the same emitter for different angles. Figure \ref{Fig: Single NV}(c) shows a polar plot of intensity as a function of the polarization analysis angle for two perpendicular polarizations. We simulated two crossed dipoles of NV center, covered in PMMA. Based on this simulation, Figure \ref{Fig: Single NV}(d) presents a map of degree of polarization ($\delta $) and expected angle ($\theta $) of the NV center crossed linear dipoles.

Figure \ref{Fig: PL} examines the PL directionality and collection efficiency of the device, analyzing how emitted photons are distributed and efficiently collected. To investigate this, we measured the PL directionality emitted from the device using back focal plane imaging, which provides a 2D angular PL intensity distribution \( I(\theta, \phi) \), where \(\theta\) and \(\phi\) are the polar and azimuthal angles, respectively (Figure \ref{Fig: PL}(a)) \cite{Abudayyeh2021QD_Placement, Lubotzky2024}.
Figure \ref{Fig: PL}(b) shows a spectrally resolved 1D cross section of the back focal plane of the device PL, capturing the broad spectral range of 620-720 nm, which includes the zero-phonon line of NV centers at 637 nm (marked with a dashed green line). The data indicates a broadband narrow angular emission pattern around 700 nm. Additionally, the PL emission over the entire measured spectrum is largely confined within 10 degrees from the normal, highlighting the broadband operational capability of the nanoantenna. Figures \ref{Fig: PL}(c) and \ref{Fig: PL}(d) display the spectrally integrated back focal plane images of the angular PL intensity distribution \( I(\theta, \phi) \) for the measured wavelength range of 620-720 nm, illustrating the highly directional nature of the emitted photons.

The simulation visualized in Figure \ref{Fig: PL}(d), using an FDTD numerical solution of a broadband optical dipole positioned in such a nanoantenna, with the experimental parameters of the actual device. We incorporated the NV spectrum measured in Figure \ref{Fig: Single ND}(e), using the measured spectral intensities as weights for the simulation data of each wavelength, and then incoherently sum the simulated intensities of all wavelengths. Additionally, the dipole orientation was fixed at \(\theta = 42^\circ\), based on the experimental estimation from the angular emission pattern, as detailed in Figures \ref{Fig: Single NV}(c) and \ref{Fig: Single NV}(d).

The CE of the emitted light was analyzed as a function of the NA of the collection optics. This was done by integrating the experimental and simulated signal over all azimuthal angles \(\phi\) within a collection cone defined by a given NA, using:

\begin{equation}
 \eta = \frac{\int_0^{2\pi}d\phi\int_0^{\theta_{NA}} d\theta \sin(\theta) I(\theta,\phi)}{\int_0^{2\pi}d\phi\int_0^{\frac{\pi}{2}}d\theta \sin(\theta) I(\theta,\phi)}.
\end{equation}

This equation, as detailed in previous studies, provides a measure of the proportion of emitted photons that can be collected into a given NA \cite{Lubotzky2024, Abudayyeh2021QD_Placement}.

Figure \ref{Fig: PL}(e) presents the CE as a function of NA for our device. The results demonstrate a CE of approximately 80\% at NA=0.5, indicating a very high directionality of the emitted photons. This performance is compared to FDTD simulation results, which show good agreement with the experimental data. In contrast, a simulation of a reference NV center placed on a glass slide (indicated by the black dashed line) shows significantly lower CE, underscoring the superior performance of our nanoantenna design.
We note that the very high accuracy $(<5 nm)$ \cite{Nikolay2018AccurateStructures} positioning method we use, allows near-optimal utilization of the nanoantenna capabilities, resulting in high CE for a broad spectrum emission into low NA optics. This allows the use of very simple collection optics and even direct coupling into an optical fiber \cite{Lubotzky2024}.

\bigskip
In summary, we demonstrated a hybrid metal-dielectric nanoantenna platform combined with ultra-precise deterministic positioning to significantly enhance photon collection from NV centers in nanodiamonds. Our approach achieves a measured CE of 82\% for NA = 0.5 and CE that approaches unity for NA$>$0.8 at room temperature. 

This work highlights the importance of optimizing both the position of the emitter by precise nano-placement and the antenna performance by a hybrid design, enabling highly directional emission while minimizing optical losses. The combination of broadband operation, high efficiency, polarization independence,  and room-temperature compatibility is essential when utilizing NV centers in integrated quantum devices, such as for single-shot spin readout or high-rate entanglement distribution in quantum networks.

\newpage


\clearpage
\section*{Acknowledgments}
NN and OB acknowledge funding by BMBF 16KISQ003 (QR.X).

\bibliography{MyCollectionB4} 

\providecommand{\latin}[1]{#1}
\makeatletter
\providecommand{\doi}
  {\begingroup\let\do\@makeother\dospecials
  \catcode`\{=1 \catcode`\}=2 \doi@aux}
\providecommand{\doi@aux}[1]{\endgroup\texttt{#1}}
\makeatother
\providecommand*\mcitethebibliography{\thebibliography}
\csname @ifundefined\endcsname{endmcitethebibliography}  {\let\endmcitethebibliography\endthebibliography}{}
\begin{mcitethebibliography}{40}
\providecommand*\natexlab[1]{#1}
\providecommand*\mciteSetBstSublistMode[1]{}
\providecommand*\mciteSetBstMaxWidthForm[2]{}
\providecommand*\mciteBstWouldAddEndPuncttrue
  {\def\EndOfBibitem{\unskip.}}
\providecommand*\mciteBstWouldAddEndPunctfalse
  {\let\EndOfBibitem\relax}
\providecommand*\mciteSetBstMidEndSepPunct[3]{}
\providecommand*\mciteSetBstSublistLabelBeginEnd[3]{}
\providecommand*\EndOfBibitem{}
\mciteSetBstSublistMode{f}
\mciteSetBstMaxWidthForm{subitem}{(\alph{mcitesubitemcount})}
\mciteSetBstSublistLabelBeginEnd
  {\mcitemaxwidthsubitemform\space}
  {\relax}
  {\relax}

\bibitem[Wolf \latin{et~al.}(2015)Wolf, Rosenberg, Rapaport, and Bar-Gill]{Wolf2015Purcell-enhancedDiamond}
Wolf,~S.~A.; Rosenberg,~I.; Rapaport,~R.; Bar-Gill,~N. {Purcell-enhanced optical spin readout of nitrogen-vacancy centers in diamond}. \emph{Physical Review B - Condensed Matter and Materials Physics} \textbf{2015}, \emph{92}, 1--5\relax
\mciteBstWouldAddEndPuncttrue
\mciteSetBstMidEndSepPunct{\mcitedefaultmidpunct}
{\mcitedefaultendpunct}{\mcitedefaultseppunct}\relax
\EndOfBibitem
\bibitem[Doherty \latin{et~al.}(2013)Doherty, Manson, Delaney, Jelezko, Wrachtrup, and Hollenberg]{Doherty2013}
Doherty,~M.~W.; Manson,~N.~B.; Delaney,~P.; Jelezko,~F.; Wrachtrup,~J.; Hollenberg,~L. C.~L. {The nitrogen-vacancy colour centre in diamond}. \emph{Physics Reports} \textbf{2013}, \emph{528}, 1--45\relax
\mciteBstWouldAddEndPuncttrue
\mciteSetBstMidEndSepPunct{\mcitedefaultmidpunct}
{\mcitedefaultendpunct}{\mcitedefaultseppunct}\relax
\EndOfBibitem
\bibitem[Guo(2023)]{Guo2023}
Guo,~S. An Overview of NV Centers. \emph{Journal of Applied Mathematics and Physics} \textbf{2023}, \emph{11}, 3666--3675\relax
\mciteBstWouldAddEndPuncttrue
\mciteSetBstMidEndSepPunct{\mcitedefaultmidpunct}
{\mcitedefaultendpunct}{\mcitedefaultseppunct}\relax
\EndOfBibitem
\bibitem[Hausmann \latin{et~al.}(2012)Hausmann, Shields, Quan, Maletinsky, McCutcheon, Choy, Babinec, Kubanek, Yacoby, Lukin, and Lončar]{Hausmann2012}
Hausmann,~B. J.~M.; Shields,~B.; Quan,~Q.; Maletinsky,~P.; McCutcheon,~M.; Choy,~J.~T.; Babinec,~T.~M.; Kubanek,~A.; Yacoby,~A.; Lukin,~M.~D.; Lončar,~M. Integrated Diamond Networks for Quantum Nanophotonics. \emph{Nano Letters} \textbf{2012}, \emph{12}, 1578--1582\relax
\mciteBstWouldAddEndPuncttrue
\mciteSetBstMidEndSepPunct{\mcitedefaultmidpunct}
{\mcitedefaultendpunct}{\mcitedefaultseppunct}\relax
\EndOfBibitem
\bibitem[Li \latin{et~al.}(2023)Li, Gerritsma, Kurdi, Codreanu, Gröblacher, Hanson, Norte, and van~der Sar]{Yufan2015}
Li,~Y.; Gerritsma,~F.~A.; Kurdi,~S.; Codreanu,~N.; Gröblacher,~S.; Hanson,~R.; Norte,~R.; van~der Sar,~T. A Fiber-Coupled Scanning Magnetometer with Nitrogen-Vacancy Spins in a Diamond Nanobeam. \emph{ACS Photonics} \textbf{2023}, \emph{10}, 1859--1865\relax
\mciteBstWouldAddEndPuncttrue
\mciteSetBstMidEndSepPunct{\mcitedefaultmidpunct}
{\mcitedefaultendpunct}{\mcitedefaultseppunct}\relax
\EndOfBibitem
\bibitem[Dhomkar \latin{et~al.}(2024)Dhomkar, Ji, Kim, Barry, and Walsworth]{Dhomkar2024}
Dhomkar,~S.; Ji,~W.; Kim,~D.; Barry,~J.~F.; Walsworth,~R.~L. Efficient readout of nitrogen-vacancy spin qubits in diamond. \emph{Science Advances} \textbf{2024}, \emph{10}, eadp6442\relax
\mciteBstWouldAddEndPuncttrue
\mciteSetBstMidEndSepPunct{\mcitedefaultmidpunct}
{\mcitedefaultendpunct}{\mcitedefaultseppunct}\relax
\EndOfBibitem
\bibitem[Robledo \latin{et~al.}(2011)Robledo, Childress, Bernien, Hanson, Hensen, and Taminiau]{Robledo2011}
Robledo,~L.; Childress,~L.; Bernien,~H.; Hanson,~R.; Hensen,~B.; Taminiau,~T.~H. High-fidelity projective read-out of a solid-state spin quantum register. \emph{Nature} \textbf{2011}, \emph{477}, 574--578\relax
\mciteBstWouldAddEndPuncttrue
\mciteSetBstMidEndSepPunct{\mcitedefaultmidpunct}
{\mcitedefaultendpunct}{\mcitedefaultseppunct}\relax
\EndOfBibitem
\bibitem[Hensen \latin{et~al.}(2015)Hensen, Bernien, Dréau, Reiserer, Kalb, Blok, Ruitenberg, Vermeulen, Schouten, Abellán, Amaya, Pruneri, Mitchell, Markham, Twitchen, Elkouss, Wehner, Taminiau, and Hanson]{Hensen2015}
Hensen,~B. \latin{et~al.}  Loophole-free Bell inequality violation using electron spins separated by 1.3 kilometres. \emph{Nature} \textbf{2015}, \emph{526}, 682--686\relax
\mciteBstWouldAddEndPuncttrue
\mciteSetBstMidEndSepPunct{\mcitedefaultmidpunct}
{\mcitedefaultendpunct}{\mcitedefaultseppunct}\relax
\EndOfBibitem
\bibitem[Hopper \latin{et~al.}(2018)Hopper, Shulevitz, and Bassett]{Hopper2018}
Hopper,~D.~A.; Shulevitz,~H.~J.; Bassett,~L.~C. Spin Readout Techniques of the Nitrogen-Vacancy Center in Diamond. \emph{Micromachines} \textbf{2018}, \emph{9}, 437\relax
\mciteBstWouldAddEndPuncttrue
\mciteSetBstMidEndSepPunct{\mcitedefaultmidpunct}
{\mcitedefaultendpunct}{\mcitedefaultseppunct}\relax
\EndOfBibitem
\bibitem[Zhang \latin{et~al.}(2021)Zhang, Guo, Ji, Wang, Yin, Kong, Lin, Yin, Shi, Wang, and Du]{Zhang2021}
Zhang,~Q.; Guo,~Y.; Ji,~W.; Wang,~M.; Yin,~J.; Kong,~F.; Lin,~Y.; Yin,~C.; Shi,~F.; Wang,~Y.; Du,~J. High-fidelity single-shot readout of single electron spin in diamond with spin-to-charge conversion. \emph{Nature Communications} \textbf{2021}, \emph{12}, 1529\relax
\mciteBstWouldAddEndPuncttrue
\mciteSetBstMidEndSepPunct{\mcitedefaultmidpunct}
{\mcitedefaultendpunct}{\mcitedefaultseppunct}\relax
\EndOfBibitem
\bibitem[Nepal \latin{et~al.}(2013)Nepal, Drummy, Biswas, Park, and Vaia]{Nepal2013LargeFluorescence}
Nepal,~D.; Drummy,~L.~F.; Biswas,~S.; Park,~K.; Vaia,~R.~A. {Large scale solution assembly of quantum dot-gold nanorod architectures with plasmon enhanced fluorescence}. \emph{ACS Nano} \textbf{2013}, \emph{7}, 9064--9074\relax
\mciteBstWouldAddEndPuncttrue
\mciteSetBstMidEndSepPunct{\mcitedefaultmidpunct}
{\mcitedefaultendpunct}{\mcitedefaultseppunct}\relax
\EndOfBibitem
\bibitem[Hoang \latin{et~al.}(2015)Hoang, Akselrod, Argyropoulos, Huang, Smith, and Mikkelsen]{Hoang2015UltrafastNanoantennas}
Hoang,~T.~B.; Akselrod,~G.~M.; Argyropoulos,~C.; Huang,~J.; Smith,~D.~R.; Mikkelsen,~M.~H. {Ultrafast spontaneous emission source using plasmonic nanoantennas}. \emph{Nature Communications} \textbf{2015}, \emph{6}\relax
\mciteBstWouldAddEndPuncttrue
\mciteSetBstMidEndSepPunct{\mcitedefaultmidpunct}
{\mcitedefaultendpunct}{\mcitedefaultseppunct}\relax
\EndOfBibitem
\bibitem[Ahmed \latin{et~al.}(2012)Ahmed, Cha, Park, Park, Lee, and Lee]{Ahmed2012PhotoluminescenceNanoneedles}
Ahmed,~S.~R.; Cha,~H.~R.; Park,~J.~Y.; Park,~E.~Y.; Lee,~D.; Lee,~J. {Photoluminescence enhancement of quantum dots on Ag nanoneedles}. \emph{Nanoscale Research Letters} \textbf{2012}, \emph{7}, 1--7\relax
\mciteBstWouldAddEndPuncttrue
\mciteSetBstMidEndSepPunct{\mcitedefaultmidpunct}
{\mcitedefaultendpunct}{\mcitedefaultseppunct}\relax
\EndOfBibitem
\bibitem[Guo \latin{et~al.}(2015)Guo, Derom, V{\"{a}}kev{\"{a}}inen, van Dijk-Moes, Liljeroth, Vanmaekelbergh, and T{\"{o}}rm{\"{a}}]{Guo2015ControllingNanoarrays}
Guo,~R.; Derom,~S.; V{\"{a}}kev{\"{a}}inen,~A.~I.; van Dijk-Moes,~R. J.~A.; Liljeroth,~P.; Vanmaekelbergh,~D.; T{\"{o}}rm{\"{a}},~P. {Controlling quantum dot emission by plasmonic nanoarrays}. \emph{Optics Express} \textbf{2015}, \emph{23}, 28206\relax
\mciteBstWouldAddEndPuncttrue
\mciteSetBstMidEndSepPunct{\mcitedefaultmidpunct}
{\mcitedefaultendpunct}{\mcitedefaultseppunct}\relax
\EndOfBibitem
\bibitem[Bitton \latin{et~al.}(2019)Bitton, Gupta, and Haran]{Bitton2019QuantumCoupling}
Bitton,~O.; Gupta,~S.~N.; Haran,~G. {Quantum dot plasmonics: From weak to strong coupling}. \emph{Nanophotonics} \textbf{2019}, \emph{8}, 559--575\relax
\mciteBstWouldAddEndPuncttrue
\mciteSetBstMidEndSepPunct{\mcitedefaultmidpunct}
{\mcitedefaultendpunct}{\mcitedefaultseppunct}\relax
\EndOfBibitem
\bibitem[Skolnick \latin{et~al.}(1998)Skolnick, Fisher, and Whittaker]{Skolnick1998StrongStructures}
Skolnick,~M.~S.; Fisher,~T.~A.; Whittaker,~D.~M. {Strong coupling phenomena in quantum microcavity structures}. \emph{Semiconductor Science and Technology} \textbf{1998}, \emph{13}, 645--669\relax
\mciteBstWouldAddEndPuncttrue
\mciteSetBstMidEndSepPunct{\mcitedefaultmidpunct}
{\mcitedefaultendpunct}{\mcitedefaultseppunct}\relax
\EndOfBibitem
\bibitem[Press \latin{et~al.}(2007)Press, G{\"{o}}tzinger, Reitzenstein, Hofmann, L{\"{o}}ffler, Kamp, Forchel, and Yamamoto]{Press2007PhotonRegime}
Press,~D.; G{\"{o}}tzinger,~S.; Reitzenstein,~S.; Hofmann,~C.; L{\"{o}}ffler,~A.; Kamp,~M.; Forchel,~A.; Yamamoto,~Y. {Photon antibunching from a single quantum-dot-microcavity system in the strong coupling regime}. \emph{Physical Review Letters} \textbf{2007}, \emph{98}, 1--4\relax
\mciteBstWouldAddEndPuncttrue
\mciteSetBstMidEndSepPunct{\mcitedefaultmidpunct}
{\mcitedefaultendpunct}{\mcitedefaultseppunct}\relax
\EndOfBibitem
\bibitem[Bogdanovi{\'{c}} \latin{et~al.}(2017)Bogdanovi{\'{c}}, {Van Dam}, Bonato, Coenen, Zwerver, Hensen, Liddy, Fink, Reiserer, Lon{\v{c}}ar, and Hanson]{Bogdanovic2017DesignNetworks}
Bogdanovi{\'{c}},~S.; {Van Dam},~S.~B.; Bonato,~C.; Coenen,~L.~C.; Zwerver,~A. M.~J.; Hensen,~B.; Liddy,~M. S.~Z.; Fink,~T.; Reiserer,~A.; Lon{\v{c}}ar,~M.; Hanson,~R. {Design and low-temperature characterization of a tunable microcavity for diamond-based quantum networks}. \emph{Applied Physics Letters} \textbf{2017}, \emph{110}\relax
\mciteBstWouldAddEndPuncttrue
\mciteSetBstMidEndSepPunct{\mcitedefaultmidpunct}
{\mcitedefaultendpunct}{\mcitedefaultseppunct}\relax
\EndOfBibitem
\bibitem[Jun \latin{et~al.}(2011)Jun, Huang, and Brongersma]{Jun2011PlasmonicEmission}
Jun,~Y.~C.; Huang,~K. C.~Y.; Brongersma,~M.~L. {Plasmonic beaming and active control over fluorescent emission}. \emph{Nature Communications} \textbf{2011}, \emph{2}\relax
\mciteBstWouldAddEndPuncttrue
\mciteSetBstMidEndSepPunct{\mcitedefaultmidpunct}
{\mcitedefaultendpunct}{\mcitedefaultseppunct}\relax
\EndOfBibitem
\bibitem[Livneh \latin{et~al.}(2015)Livneh, Harats, Yochelis, Paltiel, and Rapaport]{Livneh2015EfficientNanoantenna}
Livneh,~N.; Harats,~M.~G.; Yochelis,~S.; Paltiel,~Y.; Rapaport,~R. {Efficient Collection of Light from Colloidal Quantum Dots with a Hybrid Metal-Dielectric Nanoantenna}. \emph{ACS Photonics} \textbf{2015}, \emph{2}, 1669--1674\relax
\mciteBstWouldAddEndPuncttrue
\mciteSetBstMidEndSepPunct{\mcitedefaultmidpunct}
{\mcitedefaultendpunct}{\mcitedefaultseppunct}\relax
\EndOfBibitem
\bibitem[Livneh \latin{et~al.}(2016)Livneh, Harats, Istrati, Eisenberg, and Rapaport]{livneh2016highly}
Livneh,~N.; Harats,~M.~G.; Istrati,~D.; Eisenberg,~H.~S.; Rapaport,~R. Highly Directional Room-Temperature Single Photon Device. \emph{Nano Letters} \textbf{2016}, \emph{16}, 2527--2532\relax
\mciteBstWouldAddEndPuncttrue
\mciteSetBstMidEndSepPunct{\mcitedefaultmidpunct}
{\mcitedefaultendpunct}{\mcitedefaultseppunct}\relax
\EndOfBibitem
\bibitem[Abudayyeh and Rapaport(2017)Abudayyeh, and Rapaport]{abudayyeh2017quantum}
Abudayyeh,~H.; Rapaport,~R. Quantum Emitters Coupled to Circular Nanoantennas for High Brightness Quantum Light Sources. \emph{Quantum Science and Technology} \textbf{2017}, \emph{2}, 034004\relax
\mciteBstWouldAddEndPuncttrue
\mciteSetBstMidEndSepPunct{\mcitedefaultmidpunct}
{\mcitedefaultendpunct}{\mcitedefaultseppunct}\relax
\EndOfBibitem
\bibitem[Abudayyeh \latin{et~al.}(2021)Abudayyeh, Lubotzky, Blake, Wang, Majumder, Hu, Kim, Htoon, Bose, Malko, Hollingsworth, and Rapaport]{Abudayyeh2021QD_Placement}
Abudayyeh,~H.; Lubotzky,~B.; Blake,~A.; Wang,~J.; Majumder,~S.; Hu,~Z.; Kim,~Y.; Htoon,~H.; Bose,~R.; Malko,~A.~V.; Hollingsworth,~J.~A.; Rapaport,~R. Single photon sources with near unity collection efficiencies by deterministic placement of quantum dots in nanoantennas. \emph{APL Photonics} \textbf{2021}, \emph{6}, 036109\relax
\mciteBstWouldAddEndPuncttrue
\mciteSetBstMidEndSepPunct{\mcitedefaultmidpunct}
{\mcitedefaultendpunct}{\mcitedefaultseppunct}\relax
\EndOfBibitem
\bibitem[Lubotzky \latin{et~al.}(2024)Lubotzky, Nazarov, Abudayyeh, Antoniuk, Lettner, Agafonov, Bennett, Majumder, Chandrasekaran, Bowes, Htoon, Hollingsworth, Kubanek, and Rapaport]{Lubotzky2024}
Lubotzky,~B.; Nazarov,~A.; Abudayyeh,~H.; Antoniuk,~L.; Lettner,~N.; Agafonov,~V.; Bennett,~A.~V.; Majumder,~S.; Chandrasekaran,~V.; Bowes,~E.~G.; Htoon,~H.; Hollingsworth,~J.~A.; Kubanek,~A.; Rapaport,~R. Room-Temperature Fiber-Coupled Single-Photon Sources based on Colloidal Quantum Dots and SiV Centers in Back-Excited Nanoantennas. \emph{Nano Letters} \textbf{2024}, \emph{24}, 640--648\relax
\mciteBstWouldAddEndPuncttrue
\mciteSetBstMidEndSepPunct{\mcitedefaultmidpunct}
{\mcitedefaultendpunct}{\mcitedefaultseppunct}\relax
\EndOfBibitem
\bibitem[Waltrich \latin{et~al.}(2021)Waltrich, Lubotzky, Abudayyeh, Steiger, Fehler, Lettner, Davydov, Agafonov, Rapaport, and Kubanek]{Waltrich2021}
Waltrich,~R.; Lubotzky,~B.; Abudayyeh,~H.; Steiger,~E.~S.; Fehler,~K.~G.; Lettner,~N.; Davydov,~V.~A.; Agafonov,~V.~N.; Rapaport,~R.; Kubanek,~A. High-purity single photons obtained with moderate-NA optics from SiV center in nanodiamonds on a bullseye antenna. \emph{New Journal of Physics} \textbf{2021}, \emph{23}, 113022\relax
\mciteBstWouldAddEndPuncttrue
\mciteSetBstMidEndSepPunct{\mcitedefaultmidpunct}
{\mcitedefaultendpunct}{\mcitedefaultseppunct}\relax
\EndOfBibitem
\bibitem[Sipahigil \latin{et~al.}(2016)Sipahigil, Evans, Sukachev, Burek, Borregaard, Bhaskar, Nguyen, Pacheco, Atikian, Meuwly, Camacho, Jelezko, Bielejec, Park, Lon{\v{c}}ar, and Lukin]{Sipahigil2016AnNetworks}
Sipahigil,~A. \latin{et~al.}  {An integrated diamond nanophotonics platform for quantum-optical networks}. \emph{Science} \textbf{2016}, \emph{354}, 847--850\relax
\mciteBstWouldAddEndPuncttrue
\mciteSetBstMidEndSepPunct{\mcitedefaultmidpunct}
{\mcitedefaultendpunct}{\mcitedefaultseppunct}\relax
\EndOfBibitem
\bibitem[Gschrey \latin{et~al.}(2013)Gschrey, Gericke, Sch{\"{u}}{\ss}ler, Schmidt, Schulze, Heindel, Rodt, Strittmatter, and Reitzenstein]{Gschrey2013InSpectroscopy}
Gschrey,~M.; Gericke,~F.; Sch{\"{u}}{\ss}ler,~A.; Schmidt,~R.; Schulze,~J.~H.; Heindel,~T.; Rodt,~S.; Strittmatter,~A.; Reitzenstein,~S. {In situ electron-beam lithography of deterministic single-quantum-dot mesa-structures using low-temperature cathodoluminescence spectroscopy}. \emph{Applied Physics Letters} \textbf{2013}, \emph{102}\relax
\mciteBstWouldAddEndPuncttrue
\mciteSetBstMidEndSepPunct{\mcitedefaultmidpunct}
{\mcitedefaultendpunct}{\mcitedefaultseppunct}\relax
\EndOfBibitem
\bibitem[Shi \latin{et~al.}(2016)Shi, Sontheimer, Nikolay, Schell, Fischer, Naber, Benson, and Wegener]{Shi2016WiringLithography}
Shi,~Q.; Sontheimer,~B.; Nikolay,~N.; Schell,~A.~W.; Fischer,~J.; Naber,~A.; Benson,~O.; Wegener,~M. {Wiring up pre-characterized single-photon emitters by laser lithography}. \emph{Scientific Reports} \textbf{2016}, \emph{6}, 1--7\relax
\mciteBstWouldAddEndPuncttrue
\mciteSetBstMidEndSepPunct{\mcitedefaultmidpunct}
{\mcitedefaultendpunct}{\mcitedefaultseppunct}\relax
\EndOfBibitem
\bibitem[Harats \latin{et~al.}(2017)Harats, Livneh, and Rapaport]{Harats2017DesignEmission}
Harats,~M.~G.; Livneh,~N.; Rapaport,~R. {Design, fabrication and characterization of a hybrid metal-dielectric nanoantenna with a single nanocrystal for directional single photon emission}. \emph{Optical Materials Express} \textbf{2017}, \emph{7}, 834\relax
\mciteBstWouldAddEndPuncttrue
\mciteSetBstMidEndSepPunct{\mcitedefaultmidpunct}
{\mcitedefaultendpunct}{\mcitedefaultseppunct}\relax
\EndOfBibitem
\bibitem[Schell \latin{et~al.}(2011)Schell, Kewes, Schr{\"{o}}der, Wolters, Aichele, and Benson]{Schell2011ADevices}
Schell,~A.~W.; Kewes,~G.; Schr{\"{o}}der,~T.; Wolters,~J.; Aichele,~T.; Benson,~O. {A scanning probe-based pick-and-place procedure for assembly of integrated quantum optical hybrid devices}. \emph{Review of Scientific Instruments} \textbf{2011}, \emph{82}\relax
\mciteBstWouldAddEndPuncttrue
\mciteSetBstMidEndSepPunct{\mcitedefaultmidpunct}
{\mcitedefaultendpunct}{\mcitedefaultseppunct}\relax
\EndOfBibitem
\bibitem[Cuche \latin{et~al.}(2009)Cuche, Sonnefraud, Faklaris, Garrot, Boudou, Sauvage, Roch, Treussart, and Huant]{Cuche2009DiamondOptics}
Cuche,~A.; Sonnefraud,~Y.; Faklaris,~O.; Garrot,~D.; Boudou,~J.~P.; Sauvage,~T.; Roch,~J.~F.; Treussart,~F.; Huant,~S. {Diamond nanoparticles as photoluminescent nanoprobes for biology and near-field optics}. \emph{Journal of Luminescence} \textbf{2009}, \emph{129}, 1475--1477\relax
\mciteBstWouldAddEndPuncttrue
\mciteSetBstMidEndSepPunct{\mcitedefaultmidpunct}
{\mcitedefaultendpunct}{\mcitedefaultseppunct}\relax
\EndOfBibitem
\bibitem[{Van Der Sar} \latin{et~al.}(2009){Van Der Sar}, Heeres, Dmochowski, {De Lange}, Robledo, Oosterkamp, and Hanson]{VanDerSar2009NanopositioningCenter}
{Van Der Sar},~T.; Heeres,~E.~C.; Dmochowski,~G.~M.; {De Lange},~G.; Robledo,~L.; Oosterkamp,~T.~H.; Hanson,~R. {Nanopositioning of a diamond nanocrystal containing a single nitrogen-vacancy defect center}. \emph{Applied Physics Letters} \textbf{2009}, \emph{94}, 10--13\relax
\mciteBstWouldAddEndPuncttrue
\mciteSetBstMidEndSepPunct{\mcitedefaultmidpunct}
{\mcitedefaultendpunct}{\mcitedefaultseppunct}\relax
\EndOfBibitem
\bibitem[Huck \latin{et~al.}(2011)Huck, Kumar, Shakoor, and Andersen]{Huck2011ControlledNanowire}
Huck,~A.; Kumar,~S.; Shakoor,~A.; Andersen,~U.~L. {Controlled coupling of a single nitrogen-vacancy center to a silver nanowire}. \emph{Physical Review Letters} \textbf{2011}, \emph{106}, 2--5\relax
\mciteBstWouldAddEndPuncttrue
\mciteSetBstMidEndSepPunct{\mcitedefaultmidpunct}
{\mcitedefaultendpunct}{\mcitedefaultseppunct}\relax
\EndOfBibitem
\bibitem[Nikolay \latin{et~al.}(2018)Nikolay, Sadzak, Dohms, Lubotzky, Abudayyeh, Rapaport, and Benson]{Nikolay2018AccurateStructures}
Nikolay,~N.; Sadzak,~N.; Dohms,~A.; Lubotzky,~B.; Abudayyeh,~H.; Rapaport,~R.; Benson,~O. Accurate placement of single nanoparticles on opaque conductive structures. \emph{Applied Physics Letters} \textbf{2018}, \emph{113}, 113107\relax
\mciteBstWouldAddEndPuncttrue
\mciteSetBstMidEndSepPunct{\mcitedefaultmidpunct}
{\mcitedefaultendpunct}{\mcitedefaultseppunct}\relax
\EndOfBibitem
\bibitem[Webb \latin{et~al.}(2020)Webb, Troise, Hansen, Achard, Brinza, Staacke, Kieschnick, Meijer, Perrier, Berg-S{\o}rensen, Huck, and Andersen]{Webb2020OptimizationSignals}
Webb,~J.~L.; Troise,~L.; Hansen,~N.~W.; Achard,~J.; Brinza,~O.; Staacke,~R.; Kieschnick,~M.; Meijer,~J.; Perrier,~J.~F.; Berg-S{\o}rensen,~K.; Huck,~A.; Andersen,~U.~L. {Optimization of a Diamond Nitrogen Vacancy Centre Magnetometer for Sensing of Biological Signals}. \emph{Frontiers in Physics} \textbf{2020}, \emph{8}, 1--12\relax
\mciteBstWouldAddEndPuncttrue
\mciteSetBstMidEndSepPunct{\mcitedefaultmidpunct}
{\mcitedefaultendpunct}{\mcitedefaultseppunct}\relax
\EndOfBibitem
\bibitem[Taylor \latin{et~al.}(2008)Taylor, Cappellaro, Childress, Jiang, Budker, Hemmer, Yacoby, Walsworth, and Lukin]{Taylor2008High-sensitivityResolution}
Taylor,~J.~M.; Cappellaro,~P.; Childress,~L.; Jiang,~L.; Budker,~D.; Hemmer,~P.~R.; Yacoby,~A.; Walsworth,~R.; Lukin,~M.~D. {High-sensitivity diamond magnetometer with nanoscale resolution}. \emph{Nature Physics} \textbf{2008}, \emph{4}, 810--816\relax
\mciteBstWouldAddEndPuncttrue
\mciteSetBstMidEndSepPunct{\mcitedefaultmidpunct}
{\mcitedefaultendpunct}{\mcitedefaultseppunct}\relax
\EndOfBibitem
\bibitem[Harats \latin{et~al.}(2017)Harats, Livneh, and Rapaport]{Harats2017}
Harats,~M.~G.; Livneh,~N.; Rapaport,~R. Design, fabrication and characterization of a hybrid metal-dielectric nanoantenna with a single nanocrystal for directional single photon emission. \emph{Optical Materials Express} \textbf{2017}, \emph{7}, 834--843\relax
\mciteBstWouldAddEndPuncttrue
\mciteSetBstMidEndSepPunct{\mcitedefaultmidpunct}
{\mcitedefaultendpunct}{\mcitedefaultseppunct}\relax
\EndOfBibitem
\bibitem[Thiering and Gali(2018)Thiering, and Gali]{Thiering2018}
Thiering,~G.; Gali,~A. {Theory of the optical spin-polarization loop of the nitrogen-vacancy center in diamond}. \emph{Physical Review B} \textbf{2018}, \emph{98}, 1--12\relax
\mciteBstWouldAddEndPuncttrue
\mciteSetBstMidEndSepPunct{\mcitedefaultmidpunct}
{\mcitedefaultendpunct}{\mcitedefaultseppunct}\relax
\EndOfBibitem
\bibitem[Lethiec \latin{et~al.}(2014)Lethiec, Laverdant, Vallon, Javaux, Dubertret, Frigerio, Schwob, Coolen, and Ma{\^{i}}tre]{Lethiec2014}
Lethiec,~C.; Laverdant,~J.; Vallon,~H.; Javaux,~C.; Dubertret,~B.; Frigerio,~J.~M.; Schwob,~C.; Coolen,~L.; Ma{\^{i}}tre,~A. {Measurement of three-dimensional dipole orientation of a single fluorescent nanoemitter by emission polarization analysis}. \emph{Physical Review X} \textbf{2014}, \emph{4}, 1--12\relax
\mciteBstWouldAddEndPuncttrue
\mciteSetBstMidEndSepPunct{\mcitedefaultmidpunct}
{\mcitedefaultendpunct}{\mcitedefaultseppunct}\relax
\EndOfBibitem
\end{mcitethebibliography}
\end{document}